\documentclass[superscriptaddress,twocolumn,showpacs,prb]{revtex4}
\usepackage{graphicx}
\usepackage{subfigure}
\usepackage{color}
\usepackage{amsmath}

\begin{document}

\title{Spatial correlations in chaotic nanoscale systems
  with spin-orbit coupling}
\author{Anh T. Ngo}
\affiliation{Department of Physics and Astronomy, and
         Nanoscale \& Quantum Phenomena Institute,
         Ohio University, Athens, Ohio 45701}

\author{Eugene H. Kim}
\affiliation{Instituto de F\'isica T\'eorica, UAM-CSIC,
     Madrid 28049, Spain}
\affiliation{Department of Physics, University of Windsor,
         Windsor, Ontario, Canada N9B 3P4}

\author{Sergio E. Ulloa}
\affiliation{Department of Physics and Astronomy, and
         Nanoscale \& Quantum Phenomena Institute,
         Ohio University, Athens, Ohio 45701}

\begin{abstract}
We investigate the statistical properties of wave functions in
chaotic nanostructures with spin-orbit coupling (SOC), focussing
in particular on spatial correlations of eigenfunctions.
Numerical results from a microscopic model are compared with
results from random matrix theory
in the crossover from the gaussian orthogonal to the gaussian
symplectic ensembles (with increasing SOC);
one- and two-point distribution functions were computed to
understand the properties of eigenfunctions in this crossover.
It is found that correlations of wave function amplitudes are
suppressed with SOC; nevertheless, eigenfunction correlations
play a more important role in the two-point distribution
function(s), compared to the case with vanishing SOC.
Experimental consequences of our results are discussed.
 \end{abstract}

 \pacs{71.70.Ej ,73.63.-b, 73.63.Kv}  
\maketitle

\section{Introduction}

Spin-orbit coupling (SOC) has the potential to make novel
electronics applications  possible,\cite{SOCapplication} as it
allows one to control the electron's spin degree of
freedom through its motion.
Most systems of interest for
such applications are nano- or mesoscopic in size, including
semiconductor quantum dots,\cite{rmtdotsreview} metallic
nanoparticles,\cite{npreview} and quantum
corrals defined on surfaces.\cite{qcreview}  The energy spectrum,
and more generally, properties of these systems are (typically)
described by random matrix theory (RMT).\cite{rmtbook,chaosbook}

In RMT, the system's properties are a consequence of
its symmetries --- in the classic Wigner-Dyson
ensembles, the key symmetries are time-reversal ($T$)
and spin-rotation ($\sigma$) invariance.\cite{rmtbook,chaosbook}
With both $T$- and $\sigma$-invariance, the system
is described by the gaussian orthogonal ensemble (GOE);
SOC breaks the $\sigma$-invariance (while preserving
$T$-invariance), driving the system to the
gaussian symplectic ensemble (GSE).
[Systems with broken $T$-invariance are described by
the gaussian unitary ensemble (GUE).]
More specifically, systems with SOC are described by random
$N \times N$ matrices (with $N$$\rightarrow$$\infty$) having
quaternion components
\begin{equation}
 H = S \otimes I_2 + i \frac{\lambda}{\sqrt{4N}}
    \sum_{j=1}^3 A_j \otimes \sigma^j \, ,
\label{goegsematrix}
\end{equation}
where $S$ is an $N$$\times$$N$ symmetric matrix, and the
$\{ A_j \}$ are $N$$\times$$N$ antisymmetric matrices;
$\{ \sigma^j \}$ are the Pauli matrices, and $I_2$ is
the 2$\times$2 identity matrix.
$\lambda$ in Eq.\ \ref{goegsematrix} is related to the
SOC of the microscopic Hamiltonian --- $\lambda$=0 in the
GOE, while $\lambda$=$\sqrt{4N}$ in the GSE.

As most nanoscale systems of interest are described by
RMT,\cite{rmtdotsreview,npreview,qcreview,rmtbook,chaosbook}
it is important to understand the regimes/behaviors which
arise with SOC and the properties in these regimes.
In this work, we consider the spatial properties of wave
functions in (two-dimensional) chaotic nanoscale systems
with SOC.
The spatial properties of wave functions often determine
the system's response to experimental probes, and are
important for devices/applications.
\cite{conductanceexp,alhassid, conductance,annphys,nucleii,
rmttheoryexp,optics1,optics2}
While other works have discussed properties/consequences of
eigenvector statistics with SOC,\cite{grains,weakSO}
here we consider the spatial properties of eigenvectors
and, in particular, how these properties evolve with the
SOC.

In what follows, we consider the properties of the Hamiltonian
\begin{equation}
 H = \frac{1}{2m} {\bf p}^2
 + \alpha~ \hat{z} \cdot \left(
   {\bf p} \times \vec{\sigma} \right)
 + V({\bf r}) \ ,
\label{startinghammy}
\end{equation}
where $V({\bf r})$ is a confining and/or disorder potential.
Results obtained via RMT are compared with those obtained
by direct simulation of Eq.~\ref{startinghammy} for a
stadium billiard.\cite{bunimovich}
To characterize the system and understand its properties,
one- and two-point distribution functions were computed in
the crossover from the GOE to the GSE
(with increasing SOC).
In particular, it is found that excellent agreement between
RMT and microscopic simulations are obtained in a ``mean-field"
description of the (GOE-GSE) crossover (see below).
A key observation from our results is that correlations
of wave function amplitudes are suppressed with SOC.\@
Interestingly, however, these correlations play a more
important role in the two-point distribution function(s),
compared to the GOE (with vanishing SOC).

The rest of the paper is organized as follows.
The description of wave function statistics in RMT and, in
particular, the description of the GOE-GSE crossover is
discussed in Sec.~\ref{RMT}.
Details of our microscopic calculations --- namely the
stadium billiard considered as well as the numerical
approach employed --- are presented in Sec.~\ref{BWM}.
Our results are presented in Sec.~\ref{Res} --- one- and two-point
distribution functions obtained via RMT are compared with numerical
results from the stadium billiard.
Finally, Sec.\ \ref{concl} contains a summary of our results as
well as remarks on experimental consequences.

\section{Wave Function Statistics in RMT} \label{RMT}

In RMT, wave function correlations are governed by
the functional probability distribution\cite{srednicki,
localgaussian}
\begin{equation}
 {\cal P}(\psi) = {\cal N} \exp\left[
  -\frac{\beta}{2} \sum_{s,s'} \hspace{-0.046in}
  \int \hspace{-0.046in} d{\bf r} d{\bf r}'
  \psi^*_s({\bf r}) K_{s,s'} ({\bf r},{\bf r}')
  \psi^{\phantom *}_{s'} ({\bf r}') \right]  \, .
\label{fieldtheory}
\end{equation}
 $K_{s,s'}({\bf r},{\bf r}')$
is the functional inverse of the two-point correlation
function $\langle \psi^*_s({\bf r})
 \psi^{\phantom *}_{s'}({\bf r}') \rangle$,
where the angular brackets $\langle \cdots \rangle$
denote an average with respect to ${\cal P}(\psi)$;
the parameter $\beta$ depends on the system's symmetries
--- $\beta$=1 ($\beta$=2) in the GOE (GUE), while
$\beta$=4 in the GSE.
[${\cal N}$ is a normalization constant.]
${\cal P}(\psi)$ is the probability that a
particular energy eigenfunction with spin-$\sigma$
is equal to the specified function
$\psi_{\sigma}({\bf r})$.

A key property of
Eqs.\ \ref{goegsematrix} and \ref{startinghammy}
is their invariance
under time-reversal; as a result, the energy levels
are two-fold degenerate --- the eigenstates
$\{ \psi({\bf r}), {\cal T} \psi({\bf r}) \}$ are
degenerate, where ${\cal T}$ is the time-reversal operator.
Explicitly,
\begin{equation}
 \psi({\bf r}) =
   \left( \begin{array}{c}
    \phi({\bf r}) \\  \chi({\bf r})
  \end{array} \right)
 ,  \ \
 {\cal T} \psi({\bf r}) =
   \left( \begin{array}{c}
    -\chi^*({\bf r}) \\  \phi^*({\bf r})
  \end{array} \right)  .
\label{wavefunction}
\end{equation}
As a consequence of this two-fold degeneracy, the wave
function amplitude probed numerically and experimentally
is $|\psi_{\sigma}({\bf r})|^2$=$|\phi({\bf r})|^2
$+$|\chi({\bf r})|^2$.
As noted above, we are interested in the regimes/behaviors
which arise with SOC --- we will not only be interested in
the GSE, but also in the crossover from the GOE to the GSE.
As such, we decompose the complex wave functions
$\phi({\bf r})$ and $\chi({\bf r})$ in Eq.~\ref{wavefunction}
into their real and imaginary parts.
Then, the wave function amplitude is parameterized as
\begin{equation}
 |\psi_{\sigma}({\bf r})|^2 = \gamma_1^2~ \phi^2_1({\bf r})
   + \gamma_2^2~ \phi^2_2({\bf r}) + \gamma_3^2~ \chi^2_1({\bf r})
   + \gamma_4^2~ \chi^2_2({\bf r}),
\label{wavefunctioncrossover}
\end{equation}
where the parameters $\{\gamma_i \}$,
which satisfy the constraint
$\gamma_1^2 + \gamma_2^2 + \gamma_3^2 + \gamma_4^2 = 1$,
characterize the crossover --- $\gamma_1$=1 with
$\gamma_i$=0 for $i \neq 1$ in the GOE, while
$\gamma_i$=$1/2$ ($i = 1 \cdots 4$) in the GSE;
in the crossover, the $\{ \gamma_i \}$ fluctuate and, hence,
physical quantities must be averaged over their distribution.

\begin{figure}[t]
\centering
\includegraphics[width=3.41in]{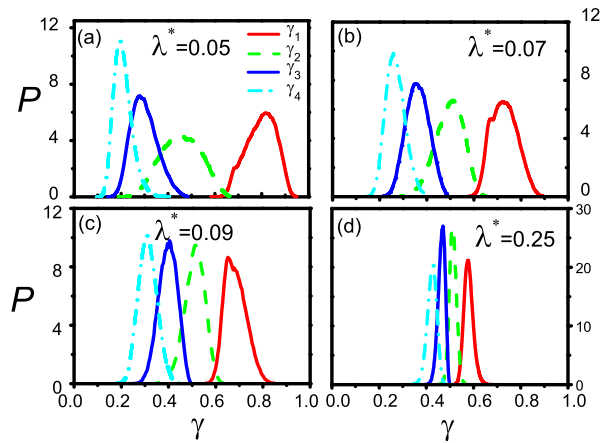}
\caption{(color online) Distribution of the $\{ \gamma_i \}$,
${\cal P}(\{ \gamma_i\})$ (from Eq.~\ref{wavefunctioncrossover}),
where $\lambda^*$=$\lambda \sqrt{4N}$.
(a) $\lambda^*=0.05$ (b) $\lambda^*=0.07$
(c) $\lambda^*=0.09$ (d) $\lambda^*=0.25$.
Notice all ${\cal P}$ change rapidly for $\lambda^* \lesssim 0.1$
and become sharply peaked at $\gamma \simeq 1/2$ for large $\lambda^*$.}
\label{fig:distribution}
\end{figure}

\begin{figure}[t]
\centering
\includegraphics[width=3.44in]{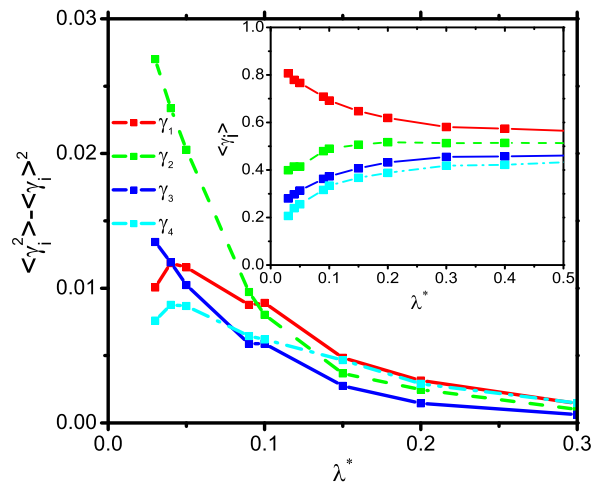}
\caption{(color online) Variance of $\{ \gamma_i \}$ vs.
$\lambda^*$=$\lambda \sqrt{4N}$.
Inset: average values of $\{ \gamma_i \}$ vs.\ $\lambda^*$.}
\label{fig:avgvariance}
\end{figure}

We obtained ${\cal P}(\{ \gamma_i \})$, the distribution of
the $\{ \gamma_i \}$, numerically from Eq.\ \ref{goegsematrix}
by considering the various orthogonal invariants\cite{goegse}
--- the results are shown in Fig.\ \ref{fig:distribution}.
We see that the ${\cal P}(\{ \gamma_i \})$ change rapidly
for $\lambda/\sqrt{4N} \lesssim 0.1$ --- in
particular, the ${\cal P}(\{ \gamma_i \})$ are broad for small
$\lambda$, but become sharply peaked gaussian-like for
larger values of $\lambda$, moving towards $\gamma_i$=1/2
($\forall~ i$) with increasing $\lambda$.
Figure \ref{fig:avgvariance} shows the variance of
the $\{ \gamma_i \}$, $\langle \gamma_i^2
\rangle$$-$$\langle \gamma_i \rangle^2$, as a function of
$\lambda$; the inset shows how the average values of the
$\{ \gamma_i \}$, $\langle \gamma_i \rangle$, evolve with
$\lambda$.
We see that the variance is extremely small for larger values
of $\lambda$; even for small values of $\lambda$ (where the
${\cal P}(\{ \gamma_i \})$ are broad and asymmetric), the
variance does not exceed 0.03.
As noted above, physical quantities must be averaged over the
${\cal P}(\{ \gamma_i \})$; however, as will be seen below,
rather good results are obtained in a ``mean-field" description
(due to the small variances), similar to what has been observed
in the GOE-GUE crossover\cite{annphys,whelan} --- rather good
results are obtained by approximating the $\{ \gamma_i \}$ by
their average values (rather than averaging over the
${\cal P}(\{ \gamma_i \}) )$.

\begin{figure}[t]
\centering
\includegraphics[height=6.1cm,width=7cm]{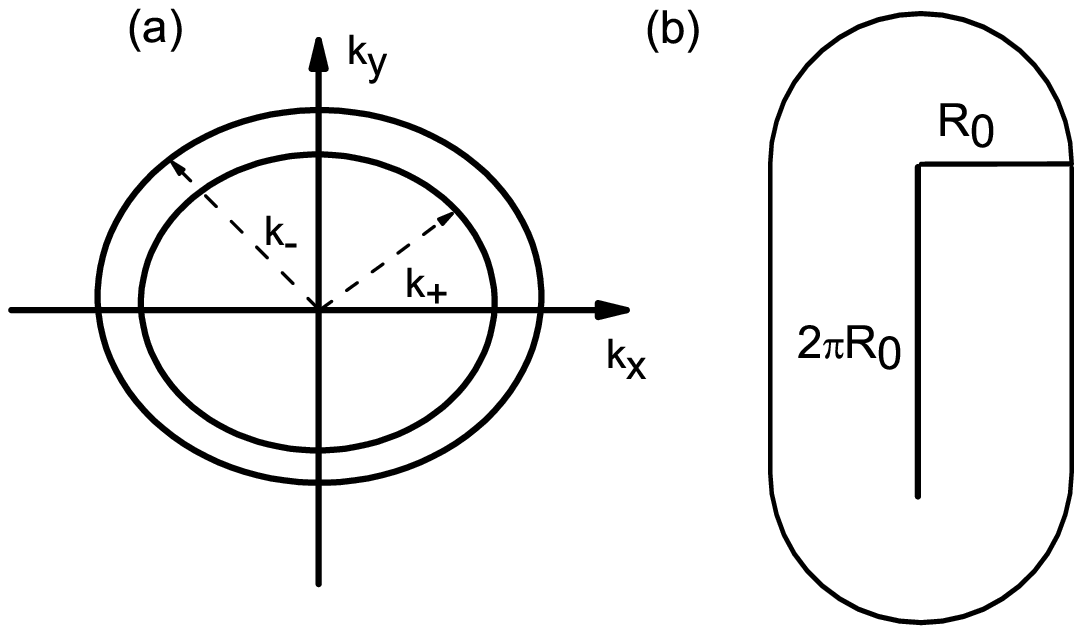}
\vspace{-0.35in} \hspace{-.11in}
\includegraphics[height=4.17cm]{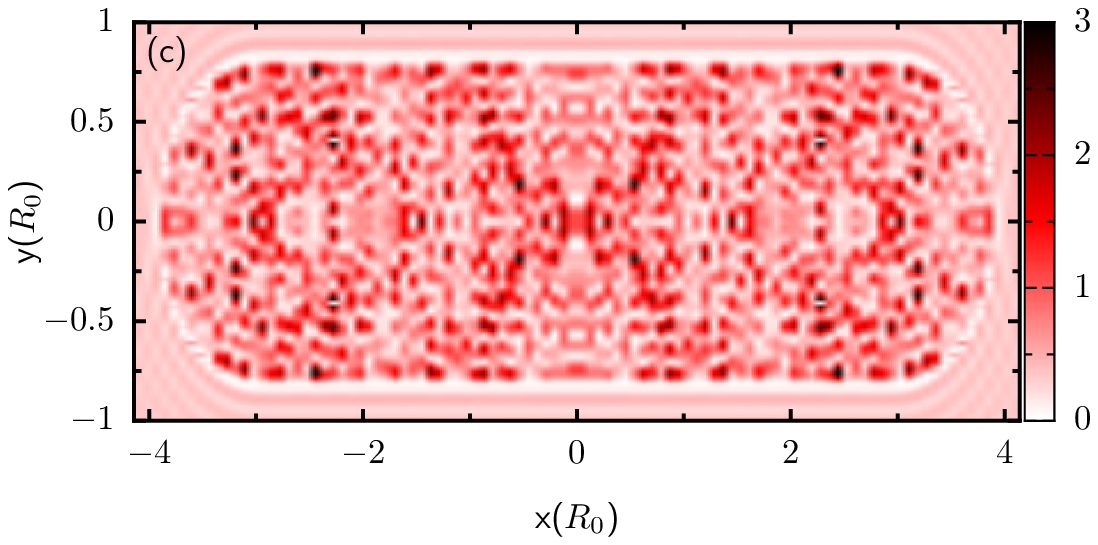}
\caption{(color online) (a) Spin-split energy surfaces with
wave vectors $k_+$ and $k_-$.
(b) Stadium billiard considered in this work.
(c) Spatial scan of the LDOS of a typical chaotic
eigenfunction.}
\label{fig:setup}
\end{figure}

From Eq.\ \ref{fieldtheory}, all spatial correlations can be
obtained once the two-point correlation function
$\langle \psi^*_s({\bf r})
 \psi^{\phantom *}_{s'}({\bf r}') \rangle$ is known.
To determine this, we expand the wave function as
\begin{equation}
 \psi({\bf r}) = \sum_{\bf p}
    \psi_{+,{\bf p}}({\bf r}) c_{+,{\bf p}}
  + \psi_{-,{\bf p}}({\bf r}) c_{-,{\bf p}}  \ ,
\label{fourier}
\end{equation}
where the two-component spinors
$\psi_{+,{\bf p}}({\bf r})$ and $\psi_{+,{\bf p}}({\bf r})$
are eigenstates of Eq.~\ref{startinghammy} with
$V({\bf r})=0$. Explicitly, the eigenvalues
$\{ E_+,E_- \}$ and corresponding eigenstates
$\{ \psi_{+,{\bf p}}({\bf r}), \psi_{-,{\bf p}}({\bf r}) \}$
are ($\hbar = 1$)
\begin{equation}
E_{\pm} = \frac{|z|^2}{2m} \pm \alpha |z|;  \
 \psi_{\pm,{\bf p}}({\bf r}) = \frac{1}{\sqrt{2A}}
   \left( \begin{array}{c}
    1 \\ \pm i z/|z|
  \end{array} \right)
  e^{i {\bf p} \cdot {\bf r}}
\nonumber
\end{equation}
where $z$=$p_x$+$i$$p_y$.
The spectrum above describes
two spin-split chiral surfaces with energy $E$, shown
schematically in Fig.~\ref{fig:setup}a,  where
$k_\pm = \sqrt{2mE+ m^2\alpha ^2} \mp m\alpha$.

To compute $\langle \psi^*_s({\bf r})
 \psi^{\phantom *}_{s'}({\bf r}') \rangle$, the Fourier
coefficients (in Eq.~\ref{fourier}) are taken to be
gaussian random variables having zero mean and variance
given by\cite{alhassid} ($a$,$b$=+,$-$)
\begin{equation}
 \langle c^*_{a, {\bf p}}
   c^{\phantom *}_{b,{\bf k}} \rangle
   = \delta_{a,b} \delta_{{\bf p},{\bf k}}
 \frac{1}{N(\epsilon)} \delta(\epsilon({\bf p}) - \epsilon)
 \, , \ \
 \langle c_{a,{\bf p}} c_{b,{\bf k}} \rangle =  0 \ .
\label{conjecture}
\end{equation}
Writing the wave function as per Eq.~\ref{wavefunction}
and using the parameterization in Eq.~\ref{wavefunctioncrossover},
we obtain\cite{translation} ($i$,$j$=1,2)
\begin{subequations}
\begin{eqnarray}
 & & \langle \phi_i({\bf r})
     \phi^{\phantom *}_{j}({\bf r}') \rangle
  = \langle \chi_i({\bf r})
     \chi^{\phantom *}_{j}({\bf r}') \rangle
  = \delta_{i,j} \, f \ ,
 \label{correlator1} \\
 & & \langle \phi_i({\bf r})
   \chi^{\phantom *}_{j}({\bf r}') \rangle
  = - \langle \chi_i({\bf r})
   \phi^{\phantom *}_{j}({\bf r}') \rangle
  = \delta_{i,j} \, g \ ,
 \label{correlator2}
\end{eqnarray}
\end{subequations}
where
\begin{subequations}
\begin{eqnarray}
 f & = & \frac{1}{2}
   \left[ J_0(k_+ R) + J_0(k_- R) \right] ,
\label{correlatorf} \\
 g & = & \frac{1}{2}
   \left[ J_1(k_+ R) - J_1(k_- R) \right]  .
\label{correlatorg}
\end{eqnarray}
\end{subequations}
In Eqs.~\ref{correlatorf} and \ref{correlatorg},
$J_0(x)$ ($J_1(x)$) is the Bessel function of order-0
(order-1),\cite{gradshteyn} $R$=$|{\bf r}-{\bf r}'|$, and
$k_{\pm}$ are the wave vectors associated with the
chiral branches at energy $E$.
The physics of Eq.~\ref{conjecture} (and Eqs.~\ref{correlator1}
and \ref{correlator2}) is that the system ergodically samples
the energy surfaces\cite{berry} (shown schematically in
Fig.\ \ref{fig:setup}a).

\section{Numerics} \label{BWM}

As described above, we are interested in comparing results
obtained via RMT with those obtained by direct simulation
of Eq.~\ref{startinghammy}. To this end, we have computed the
local density of states (LDOS) for a stadium
billiard,\cite{bunimovich} where the billiard's wall was
constructed with a unitary delta-function
potential\cite{boundarywall}
\begin{equation}
 V({\bf r}) = V_0~ \delta\left( {\bf r} - {\bf R}(s) \right) \ ,
\end{equation}
with ${\bf R}(s)$ parameterizing the wall
(and $V_0 \rightarrow \infty$).
The retarded Green's function (GF) for the system,
\begin{equation}
 G({\bf r}, {\bf r'}; \omega) = \langle {\bf r} |
   \left( \omega - H + i0^{+} \right)^{-1} | {\bf r}' \rangle ,
\end{equation}
is computed from the Dyson equation
\begin{eqnarray}
 G({\bf r}, {\bf r}'; \omega) & = &
    G_0({\bf r}, {\bf r}'; \omega)  \nonumber \\
  & + & V_0~ \int_{\cal C} ds~ G_0({\bf r}, {\bf R}(s); \omega)~
     G({\bf R}(s), {\bf r}'; \omega) \ .
\nonumber
\end{eqnarray}
In this equation, $G_0({\bf r}, {\bf r'}; \omega)$ is the
free-particle GF, i.e. the GF in the absence of the corral's
wall, but in the presence of SOC,\cite{spinorbitgreens,soccorral}
\begin{equation}
 G_{0}({\bf r},{\bf r}';\omega) = G^{0}_0(R;\omega) ~I
 + G^{1}_0(R;\omega) \left( \begin{array}{c c}
    0 & -i e^{-i\theta} \\
    i e^{i\theta} & 0 \end{array} \right)
\nonumber
\end{equation}
where
\begin{eqnarray}
 G^{0}_{0}(R;\omega) & = & -i \frac{m}{2k}
  \left\{ k_{-} H^{(1)}_0 (R k_-)
        + k_{+} H^{(1)}_0 (R k_+)  \right\}  \ ,
     \nonumber  \\
 G^{1}_{0}(R;\omega) & = & - \frac{m}{2k}
  \left\{ k_{-} H^{(1)}_1 (Rk_-)
        - k _{+} H^{(1)}_1 (Rk_+ ) \right\}   \ ,
\nonumber \\
\end{eqnarray}
and
$\exp(i\theta) = [(x-x')+i(y-y')]/R$, with
$H^{(1)}_0(x)$ and $H^{(1)}_1(x)$ being Hankel
functions,\cite{gradshteyn}
with $R$=$|{\bf r}-{\bf r}'|$ and $k_\pm$ defined as before.
The LDOS is then obtained from the GF via
$A({\bf r},\omega) = -(1/\pi) {\rm I m} \, {\rm Tr}
  \left[ G({\bf r}, {\bf r}; \omega) \right]$.

The stadium billiard we consider is shown schematically in
Fig.\ \ref{fig:setup}b. With energy in units of $E_0$=$
1/(2m R_0^2)$ and SOC in units of $\alpha_0$=1/$(mR_0)$,
where $R_0$ is the radius of the stadium's circular cap,
we have considered eigenstates with energy
$E$$\simeq$$405 E_0$, and have investigated SOCs in the
range $0$$\leq$$\alpha$$\leq 10\alpha_0$.
[Choosing $R_0$=70\AA, and $m$=$0.26m_e$ (with $m_e$ being the
electron's rest mass), one obtains
$\alpha_0$=$3.7\times10^{-11}$eVm, a value consistent with
e.g. electrons on an Au(111) surface.\cite{gold111,soccorral}]
A spatial scan of the LDOS for a typical eigenstate considered
is shown in Fig.\ \ref{fig:setup}c;
from the LDOS, one- and two-point distribution functions were
computed, going from the GOE to the GSE (with increasing $\alpha$).

\section{Results} \label{Res}

We now analyze the properties of the system, comparing
results from RMT with those obtained by direct simulation
of Eq.\ \ref{startinghammy} for a stadium billiard.
We begin by determining the regimes which arise as
function of the SOC strength.
To this end, we consider the one-point function
$ {\cal P}(\nu) = \langle \delta \left(
\nu - A |\psi_{\sigma} ({\bf r})|^2 \right) \rangle$,
which is obtained from Eq.~\ref{fieldtheory} by integrating
out the degrees of freedom except at ${\bf r}$.
Using Eq.\ \ref{wavefunctioncrossover}, we obtain
\begin{eqnarray}
 & & {\cal P}(\nu) = \frac{\nu}
 {4\gamma_1\gamma_2\gamma_3\gamma_4} \int_0^1 dz~
 \label{goegsecrossover}  \\  & & \hspace{0.087in} \times ~
 \exp\left\{-\frac{\nu}{4} \left[ (1-z)
 \left(\frac{1}{\gamma_1^2}+\frac{1}{\gamma_2^2} \right)
 + z \left(\frac{1}{\gamma_3^2}+\frac{1}{\gamma_3^2}
 \right) \right] \right\}
 \nonumber  \label{1ptfunction}
 \\ & & \hspace{0.087in} \times ~
 I_0\left[\frac{\nu}{4} \left(\frac{1}{\gamma_1^2}
  - \frac{1}{\gamma_2^2}\right) (1-z)\right]
 I_0\left[\frac{\nu}{4} \left(\frac{1}{\gamma_3^2}
  - \frac{1}{\gamma_4^2}\right) z  \right] ,
\nonumber
\end{eqnarray}
where $I_0(x)$ is the modified Bessel function of
order-zero.\cite{gradshteyn}
This expression reduces to
${\cal P}_{\rm GOE}(\nu) = \exp(-\nu/2)/\sqrt{2\pi \nu}$ in
the GOE ($\gamma_1$=1 and $\{\gamma_i\}$=0 for $i$$\neq$1)
and ${\cal P}_{\rm GSE}(\nu) = 4\nu \exp (-2 \nu)$ in
the GSE ($\gamma_i$=1/2 $\forall i$).

\begin{figure}[b]
\includegraphics[width=3.41in]{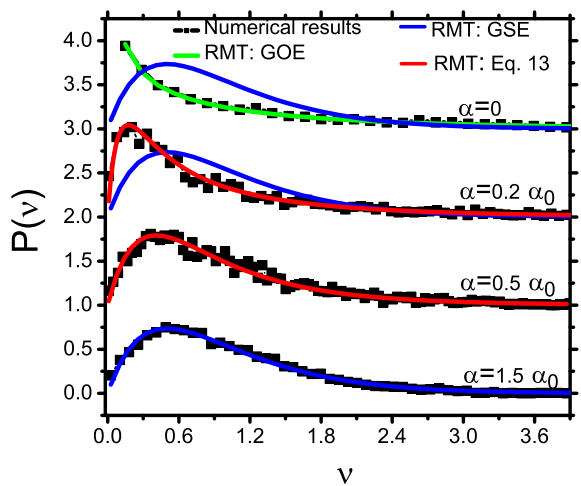}
\caption{(color online)
$ {\cal P}(\nu) = \langle \delta \left( \nu
- A |\psi_{\sigma} ({\bf r})|^2 \right) \rangle$.
From top to bottom:
$\alpha$=0 (GOE), $\alpha$=0.2,
$\alpha$=0.5, $\alpha$=1.5 (GSE).
Each curve has been vertically offset by one unit
for clarity.}
\label{fig:onept}
\end{figure}

Figure \ref{fig:onept} shows numerical results for
${\cal P}(\nu)$ for different values of the SOC; the
results are compared with Eq.~\ref{goegsecrossover}
in a ``mean-field" description, i.e.
with the $\{ \gamma_i \}$ evaluated at their average
values --- for $\alpha$=0.2$\alpha_0$ ($\alpha$=0.5$\alpha_0$),
we find $\lambda = 0.04\sqrt{4N}$
($\lambda = 0.08\sqrt{4N}$).\cite{fitting}
The physics of Eq.~\ref{startinghammy} is determined by
its two length scales --- the spin-flip length
$l_{\rm sf}$=$1/(m\alpha)$ and the linear dimension
of the system $L$ ($\simeq R_0$).
Figure \ref{fig:onept} shows how the system evolves toward
the GSE as the SOC is increased.
In particular, we find the system to be in the GSE for
$\alpha \gtrsim 1.5 \alpha_0$ i.e. $\l_{\rm sf} \lesssim 2R_0/3$;
once the system is in this GSE regime, the statistics
do not change further as the SOC is increased.

\begin{figure}[t]
\includegraphics[width=3.41in]{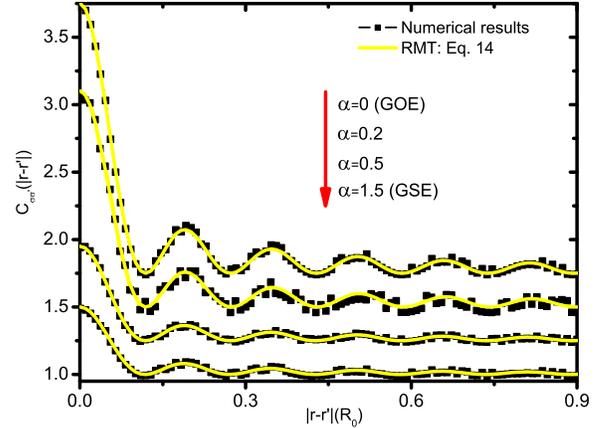}
\caption{(color online) Amplitude correlator
${\cal C}_{\sigma \sigma'}({\bf r},{\bf r}')
= \langle A |\psi_{\sigma}({\bf r})|^2
A |\psi_{\sigma '}({\bf r}')|^2 \rangle$.
From top to bottom:
$\alpha$=0 (GOE), $\alpha$=0.2$\alpha_0$,
$\alpha$=0.5$\alpha_0$, $\alpha$=1.5$\alpha_0$ (GSE).
Each curve has been vertically offset by 1/4 unit for
clarity.}
\label{fig:doscorrelate}
\end{figure}

We now turn to spatial correlations of eigenfunctions.
We first consider the amplitude correlator
${\cal C}_{\sigma \sigma'}({\bf r},{\bf r}')
= \langle A |\psi_{\sigma}({\bf r})|^2
A |\psi_{\sigma '}({\bf r}')|^2 \rangle$.
Using the parameterization in Eq.~\ref{wavefunctioncrossover},
we obtain
\begin{eqnarray}
 & & {\cal C}_{\sigma \sigma'}({\bf r},{\bf r}')
   = 1 + 2 \left[ (\gamma_1^4 + \gamma_2^4
             + \gamma_3^4 +\gamma_4^4) f^2
 \right. \nonumber  \\ & & \left. \hspace{1in}
   + 2 \left( \gamma_1^2 \gamma_3^2 + \gamma_2^2 \gamma_4^2
     \right) g^2 \right] \ .
\label{doscorrelator}
\end{eqnarray}
Notice that this reduces to
${\cal C}_{\sigma \sigma '}^{\rm GOE} ({\bf r}, {\bf r}')
= 1 + 2f^2$ in the GOE, and to
${\cal C}_{\sigma \sigma '}^{\rm GSE} ({\bf r}, {\bf r}')
= 1 + {\cal V}^2/2$ in the GSE
where ${\cal V}^2$=$f^2$+$g^2$.
Numerical results for
${\cal C}_{\sigma \sigma '} ({\bf r}, {\bf r}')$
are shown in Fig.\ \ref{fig:doscorrelate}
and are compared with Eq.~\ref{doscorrelator},
with the $\{ \gamma_i \}$ evaluated at their average
values (as before).
We see that the maximum is larger in the GOE; more generally,
the correlations decay more rapidly with SOC --- amplitude
correlations are suppressed as $\sigma$-invariance is
broken.

Having determined the parameter regimes and, in particular,
how large the SOC must be to be in the GSE, we now consider
in greater detail the properties of the system in the GSE.
To this end, we consider the joint distribution function
${\cal P}(\nu_1,\nu_2) = \left\langle
  \delta \left( \nu_1 - A | \psi_{\sigma}({\bf r}) |^2 \right)
  \delta \left( \nu_2 - A | \psi_{\sigma'}({\bf r}') |^2 \right)
  \right\rangle$,
which is obtained from Eq.~\ref{fieldtheory} by integrating
out the degrees of freedom except those at
${\bf r}$ and ${\bf r}'$.
For the GSE we obtain
\begin{equation}
 {\cal P}_{\rm GSE}(\nu_1, \nu_2) =
   \frac{8\sqrt{\nu_1 \nu_2}}{{\cal V}(1-{\cal V}^2)}
   \exp \left( -2 {\cal X}_S \right)
   I_1\left( 4 {\cal X}_P \right)  \ ,
\label{jointsymp}
\end{equation}
where $I_1(x)$ is the modified Bessel function of
order-$1$;\cite{gradshteyn}
for comparison, we also consider
${\cal P}(\nu_1,\nu_2)$ in the GOE\cite{srednicki}
\begin{equation}
 {\cal P}_{\rm GOE}(\nu_1, \nu_2) =
  \frac{\exp \left( - {\cal X}_S/2 \right)
  \cosh \left( {\cal X}_P \right)}
  {2\pi \sqrt{1-f^2} \sqrt{\nu_1 \nu_2} } \ .
\nonumber
\end{equation}
In the above equations,
${\cal X}_S$=$(\nu_1 + \nu_2)/(1-{\cal X}^2)$
and
${\cal X}_P$=${\cal X}\sqrt{\nu_1 \nu_2}/(1-{\cal X}^2)$
where ${\cal X}$=${\cal V}$ (${\cal X}$=$f$) for the GSE (GOE).

\begin{figure}[t]
\centering
\includegraphics[width=3.5in]{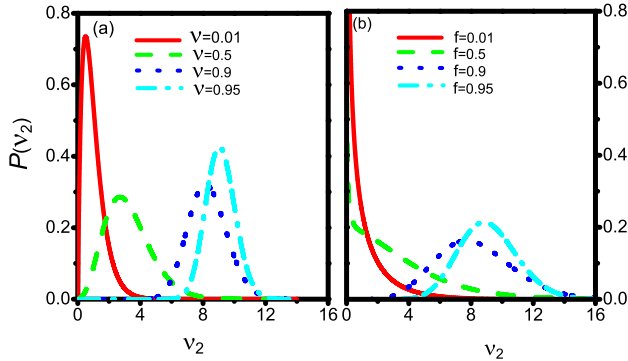}
\caption{(color online) Conditional probability
(a) ${\cal P}^{\rm GSE}_{\nu_1}({\nu_2})$, and
(b) ${\cal P}^{\rm GOE}_{\nu_1}({\nu_2})$, for
several values of ${\cal V}$ ($f$), for $\nu_1=10$.}
\label{fig:conditional}
\end{figure}

We now consider the properties and consequences of
${\cal P}({\nu_1},{\nu_2})$.
We begin by considering the conditional probability
\begin{equation}
 {\cal P}_{\nu_1}({\nu_2})
 = {\cal P}(\nu_1,\nu_2) / {\cal P}(\nu_1)
\label{conditional}
\end{equation}
which describes the  wave function distribution at ${\bf
r}_2$, provided $A |\psi({\bf r}_1)|^2 = \nu_1$.
It follows from Eq.~\ref{conditional}  that correlations
between fluctuations at different points depend on their
amplitudes\cite{prigodin} --- regions of high amplitude
(i.e. large $\nu_1$) are correlated over larger distances,
while regions of small amplitude are correlated over shorter
distances.
${\cal P}^{\rm GSE}_{\nu_1}({\nu_2})$ for the GSE is
shown in Fig.~\ref{fig:conditional}a for several values
of ${\cal V}=\sqrt{f^2+g^2}$;
${\cal P}^{\rm GOE}_{\nu_1}({\nu_2})$ for the GOE is
shown in Fig.~\ref{fig:conditional}b for comparison,
for several values of $f$.

From Eq.~\ref{conditional}, one can obtain the average
$\langle \nu_2 \rangle_{\nu_1}$ and the mean squared
fluctuation $\langle (\delta\nu_2)^2 \rangle_{\nu_1}
 = \langle \nu_2^2 \rangle^{\phantom 2}_{\nu_1}
 - \langle \nu_2 \rangle^2_{\nu_1}$,
where  $\langle \cdots \rangle_{\nu_1}$ denotes an average
with respect to $ {\cal P}_{\nu_1}({\nu_2})$:
\begin{eqnarray}
 \langle \nu_2 \rangle_{\nu_1} & = & 1
  + {\cal X}^2(\nu_1 - 1) ,  \label{conditionalavg} \\
 \langle (\delta \nu_2)^2 \rangle_{\nu_1}
  & = & {\cal C} \left[ 1 + 2{\cal X}^2(\nu_1 - 1)
  + {\cal X}^4 (1 - 2\nu_1) \right] , \nonumber
\end{eqnarray}
where ${\cal C}$=2 for the GOE,\cite{rmttheoryexp}
while ${\cal C}$=1/2 for the GSE.  [As before,
${\cal X}$=${\cal V}$ ($f$) for the GSE (GOE).]
From this, we see that fluctuations are suppressed in the
GSE compared to the GOE.
More generally, fluctuations are largest in the GOE
(compared with the GUE\cite{prigodin} and the
GSE, Eq.~\ref{conditionalavg}) and, hence,
correlations are the weakest.

We now consider the distribution of the product
$A |\psi_{\sigma}({\bf r})
 \psi_{\sigma '}({\bf r}')|$,
${\cal P}(\Gamma) = \langle \delta \left( \Gamma
  - A|\psi_{\sigma}({\bf r}) \psi_{\sigma'}({\bf r}')|
    \right) \rangle$.
${\cal P}(\Gamma)$ determines a number of experimentally
relevant quantities, such as the form factor
in resonant scattering in
complex nucleii,\cite{nucleii} amplitudes in tunneling measurements,
and the conductance amplitude distribution through small quantum
dots.\cite{conductance}
From Eq.~\ref{jointsymp}, we obtain for the GSE
\begin{equation}
 {\cal P}_{\rm GSE}(\Gamma) =
   \frac{32~ \Gamma^{2}}{|{\cal V}|(1-{\cal V}^2)}
   I_1\left( \frac{4|{\cal V}|~\Gamma}{1-{\cal V}^2}\right)
   K_0 \left( \frac{4~\Gamma}{1-{\cal V}^2} \right) \ ,
\label{proddistribute}
\end{equation}
where $K_0(x)$ is a modified Bessel functions of
order-zero;\cite{gradshteyn}
in the GOE, we obtain
\begin{equation}
 {\cal P}_{\rm GOE}(\Gamma) =
  \frac{2}{\pi \sqrt{1-f^2}}
  K_0 \left( \frac{\Gamma}{1-f^2} \right)
  \cosh \left( \frac{f~\Gamma}{1-f^2} \right) \ .
  \nonumber
\end{equation}

\begin{figure}[t]
\centering
\includegraphics[height=2.8in]{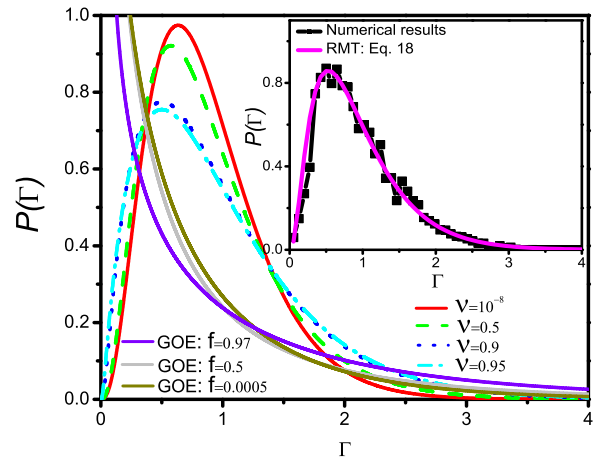}
\caption{(color online) Product distribution
${\cal P}(\Gamma) = \langle \delta \left( \Gamma
  - A|\psi_{\sigma}({\bf r}) \psi_{\sigma'}({\bf r}')|
    \right) \rangle$
in the GSE for several values of ${\cal V}$.
For comparison, ${\cal P}_{\rm GOE}(\Gamma)$ is also shown for $f=0.5$.
Inset: Comparison of numerical and RMT results for $R=0.055R_0$.}
\label{fig:twopoint}
\end{figure}

Figure \ref{fig:twopoint} shows results for ${\cal P}(\Gamma)$
for several values of ${\cal V}$ ($f$) for the GSE (GOE).
We see that the maximum of ${\cal P}_{\rm GSE}(\Gamma)$
decreases with increasing  ${\cal V}$ with the tail becoming
slightly longer.
For comparison, ${\cal P}_{\rm GOE}(\Gamma)$ is shown for
different values of $f$.
We see that correlations play a more significant role in the
GSE --- indeed, except for a very small region near $\Gamma$=0,
${\cal P}_{\rm GOE}(\Gamma)$ is essentially indistinguishable
from the result with $f$$\rightarrow$0.
This is a consequence of the fact that fluctuations are largest
in the GOE and correlations are the weakest.
Shown in the inset are numerical results for
${\cal P}_{\rm GSE}(\Gamma)$ for $R$=$0.055R_0$
in comparison with the RMT result, Eq.~\ref{proddistribute}.

\section{Concluding Remarks} \label{concl}

To summarize,
we have investigated the statistical properties of wave functions
in (two-dimensional) chaotic nanostructures with spin-orbit
interactions, focussing particularly on spatial correlations
of eigenfunctions.
Numerical results obtained for a chaotic stadium billiard
were compared with (analytic) results from RMT.
It was found that excellent agreement between RMT and microscopic
simulations are obtained in a ``mean-field" description of the
GOE-GSE crossover.
A key observation from our results is that correlations of
wave function amplitudes are suppressed with SOC.\@
Interestingly, however, these correlations with SOC play
a more significant role in the two-point distribution
function(s).

Our results have implications for a number of systems of
current interest.
Indeed, the effects of SOC have been observed in transport
through quantum dots.\cite{socdot}
These effects could also be observed in quantum corrals defined
on Au and Ag (111) surfaces,\cite{soccorral}
where large SOC has been observed recently,\cite{goldsurface}
especially as scanning tunneling microscopy techniques have
exquisite control of positioning and correlation measurements.

\section*{Acknowledgements}

We acknowledge helpful conversations with H.~U. Baranger
and P.~W. Brouwer.
EHK acknowledges the warm hospitality of the
Instituto de F\'isica T\'eorica (Madrid), where most of
this work was performed.
This work was supported in Ohio by NSF-DMR MWN/CIAM and
NSF-PIRE grants.


\begin{thebibliography}{100}
\bibitem{translation}
Eqs.~\ref{correlator1} and \ref{correlator2} were computed in an
approximation which assumes translation invariance; this is known to
give reasonable results away from the boundaries.
\bibitem{fitting}$\lambda$ (in Eq.\ \ref{goegsematrix})
was determined by minimizing the $\chi^2$ deviation between Eq.\ 13
and the numerical data.
\end{thebibliography}
\end{document}